\documentclass[utf8]{FrontiersinHarvard}
\usepackage{url,hyperref,lineno,microtype,subcaption}
\usepackage[onehalfspacing]{setspace}
\usepackage{amsmath,amssymb}
\usepackage{float}
\usepackage{placeins}
\usepackage{needspace}

\DeclareRobustCommand{\ion}[2]{\textup{#1\,\textsc{\lowercase{#2}}}}

\newcommand*\aap{A\&A}\newcommand*\apj{ApJ}\newcommand*\apjl{ApJ}\newcommand*\mnras{MNRAS}\newcommand*\nat{Nature}\newcommand*\solphys{Sol.~Phys.}\newcommand{\Icore}{I_{\rm core}}

\def\keyFont{\fontsize{8}{11}\helveticabold}\def\firstAuthorLast{Sitterson {et~al.}}\def\Authors{Taylor Sitterson$^{1,2}$, 
Jo\~ao M. da Silva Santos$^{2,3,*}$, 
Joshua Bentley$^{1}$,\\
Tom Schad$^{4}$,
Momchil Molnar$^{5}$,
Bart De Pontieu$^{6,7,8}$, 
Juan Martinez-Sykora$^{9,6,7,8}$, 
Alexander G. M. Pietrow$^{10}$
}

\def\Address{$^{1}$ Department of Astrophysical and Planetary Sciences, University of Colorado Boulder, 2000 Colorado Avenue, Boulder, CO 80309, USA
$\newline$$^{2}$ National Solar Observatory, 3665 Discovery Drive, Boulder, CO 80303, USA
$\newline$$^{3}$ Department of Physics, University of Colorado Boulder, 2000 Colorado Avenue, Boulder, CO 80309, USA
$\newline$$^{4}$ National Solar Observatory, 22 \textquotesingle Ohi\textquotesingle a K\={u} Street, Pukalani, HI 96768, USA
$\newline$$^{5}$ Southwest Research Institute,  Boulder, CO, 80302, USA
$\newline$$^{6}$ Lockheed Martin Solar and Astrophysics Laboratory, 3251 Hanover Street, Palo Alto, CA 94304, USA
$\newline$$^{7}$ Rosseland Centre for Solar Physics, University of Oslo, PO Box 1029 Blindern, 0315 Oslo, Norway
$\newline$$^{8}$ Institute of Theoretical Astrophysics, University of Oslo, PO Box 1029 Blindern, 0315 Oslo, Norway 
$\newline$$^{9}$ SETI Institute, 339 N Bernardo Ave Suite 200, Mountain View, CA 94043, USA 
$\newline$$^{10}$ Leibniz-Institut f\"ur Astrophysik Potsdam (AIP), An der Sternwarte 16, 14482 Potsdam, Germany
}

\def\corrAuthor{J. M. da Silva Santos}
\def\corrEmail{joao.daSilvaSantos@colorado.edu}

\begin{document}\onecolumn\firstpage{1}

\title[Magnetoacoustic Waves in Solar Plage]
{Multi-Height Spectropolarimetric Signatures of Magnetoacoustic Waves in Solar Plage}

\author[\firstAuthorLast]{\Authors}\address{}\correspondance{}\extraAuth{}

\maketitle

\begin{abstract}
We analyze high-cadence spectropolarimetric observations of solar plage obtained with the Visible Spectro-Polarimeter (ViSP) on the Daniel K. Inouye Solar Telescope (DKIST) to investigate low-chromospheric waves sampled by the \ion{Na}{I} D$_1$ 5896\,\AA\ and \ion{Ca}{II} 8542\,\AA\ lines. Observations were taken at an oblique viewing angle, providing sensitivity to transverse motions. Magnetic oscillations within plage elements exhibit significant line-of-sight (LOS) RMS amplitudes of $\delta B_{\rm LOS}\approx9$\,G on average in the \ion{Na}{I} D$_1$ line, whereas the corresponding \ion{Ca}{II} oscillations are only marginally above the noise level. Adaptive profile tracking provided line-core intensities and Doppler velocities, the weak-field approximation provided LOS magnetic fields, and non-LTE inversions provided mass densities and formation heights. Cross-spectral analysis shows velocity--intensity phase distributions coupled at $-90^{\circ}$, consistent with compressive oscillations. The velocity phases in  \ion{Na}{I}--\ion{Ca}{II} indicate upward propagation and the phase speeds are predominantly sub-Alfv\'enic, consistent with slow magnetoacoustic waves. Although the slit geometry prevents a definitive identification of the tube-eigenmode, the differing magnetic phase relationships suggest a height-dependent sausage-like behavior in the coupling between the magnetic and compressive perturbations.  Additionally, our inferred wave fluxes are insufficient to heat active region plage. 

\tiny\keyFont{\section{Keywords:} Sun, chromosphere, magnetic fields, magnetohydrodynamic waves, solar plage}
\end{abstract}

\section{Introduction}

Plage regions consist of strong concentrations of nearly vertical and predominantly unipolar magnetic flux rooted in the photosphere \citep[e.g.,][]{Topka1992,MartinezPillet1997,Buehler2015,2020A&A...644A..43P,2023ApJ...954L..35D, Cretignier2024}. Moreover, they show chromospheric radiative losses that not only are greatly enhanced relative to the quiet Sun in average terms \citep[e.g.,][]{2009ApJ...707..482F}, but are also spatially and temporally structured due to modified thermal and magnetic stratification \citep[e.g.,][]{2015ApJ...809L..30C,Anan2021,Morosin2022}. Determining how energy is transported to and dissipated within plage is essential to understand the thermal balance of the chromosphere at the base of hot coronal loops \citep[][and references therein]{Carlsson2019}. Magnetohydrodynamic waves are natural candidates because photospheric magnetic elements can be buffeted by convection and can support slow magnetoacoustic, kink, sausage-like, and Alfv\'enic disturbances capable of transporting energy upward \citep[e.g.,][]{Fujimura2009,Jess2009,Stangalini2013,Stangalini2014,Stangalini2017,Jess2015}. In magnetized regions, the field geometry can guide waves, modify the acoustic cutoff frequency, and influence where wave energy is reflected, transmitted, converted, and dissipated \citep[e.g.,][]{BelLeroy1977,McIntosh2006,Jefferies2006}. Other physical mechanisms relevant to atmospheric heating in plage regions are summarized in \citet{Anan2021}.

Photospheric studies have measured velocity and magnetic oscillations with dominant periods of approximately 3--5 min and have demonstrated that oscillatory power and phase relations depend on the magnetic-field strength and inclination \citep{Howard1967,Centeno2009,deWijn2009,Kostik2013,Norton2021}. Chromospheric observations have revealed propagating magnetoacoustic disturbances, shocks, and dynamic fibrils rooted in plage magnetic concentrations \citep[e.g.,][]{2003ApJ...595L..63D,Hansteen2006,Kayshap2020,Ji2021}.

Acoustic-wave energy fluxes in plage have also been compared with chromospheric radiative-loss requirements, although the conclusions vary depending on the height range considered \citep[e.g.,][]{Sobotka2016,Abbasvand2020a,Abbasvand2020b,Abbasvand2021,Molnar2021,Morosin2022,daSilvaSantos2024}. Uncertainties in formation heights, acoustic cutoffs, and mass densities can significantly affect inferred wave properties and energy fluxes \citep[][]{2023ApJ...945..154M}. Three-dimensional radiation-MHD simulations similarly indicate that slow magnetoacoustic waves can carry substantial energy upward in plages, but limited spatial resolution can underestimate wave flux \citep{Yadav2021}. Vortex flows may further enhance the dissipation of such wave modes \citep{2026arXiv260521230Y}.

The observational connection between chromospheric wave signatures, magnetic-field perturbations, and local energy transport in plage remains incomplete.  This is challenging to assess observationally because MHD wave perturbations in the magnetic field may be only a few gauss \citep[e.g.,][]{2016ApJ...819L..11S}. Other outstanding questions concern how wave properties evolve through the chromosphere, how the inferred wave-energy flux depends on the magnetic field and the spatial organization of plage flux tubes \citep[][and references therein]{Srivastava2021}, and how ion--neutral collisions contribute to wave-energy dissipation \citep[][and references therein]{2017PPCF...59a4038K}.

Here we used spectropolarimetric observations obtained with the Daniel K. Inouye Solar Telescope \citep[DKIST;][]{Rimmele2020} to investigate oscillations in a solar plage region. By combining measurements of line-core intensity, Doppler velocity, and line-of-sight (LOS) magnetic field perturbations in the \ion{Na}{I} D$_1$ 5896\,\AA\ and \ion{Ca}{II} 8542\,\AA~lines, we tested whether the observed oscillations are consistent with upward propagating slow magnetoacoustic waves, quantified the associated wave-energy fluxes, and investigated how they correlate with the underlying magnetic field. 

\section{Data \& Methods}

\subsection{Data Acquisition and Postprocessing}

We used spectropolarimetric data obtained with the Visible Spectro-Polarimeter \citep[ViSP;][]{2022SoPh..297...22D} at DKIST on 2025 July 25 during observing Cycle 3 (proposal PID\_3\_4). The target was an active-region plage in NOAA 14153, observed at $\mu=0.76$, where $\mu$ is the cosine of the heliocentric angle. To improve the coordinate information of the ViSP data, we used images in the visible continuum provided by the Helioseismic and Magnetic Imager \citep[HMI;][]{2012SoPh..275..207S} for coalignment. For context, we also used a 1600\,\AA~continuum image taken with the Atmospheric Imaging Assembly \citep[AIA;][]{2012SoPh..275...17L}. 

The observing program included both high-sensitivity, moderate-size ($\sim$12.3" in width) rasters to characterize the overall magnetic topology and narrower ($\sim$1.5" in width), high-cadence rasters for time-series analysis. In both cases, the ViSP slit was oriented to track the parallactic angle to minimize wavelength-dependent refraction perpendicular to the slit. This results in a slight FOV rotation of approximately $\sim$1.5$^\circ$ over the observing sequence, which is not expected to significantly affect the frequency range considered in this work. Here, we focus primarily on the high-cadence program, obtained between 19:17~UT and 19:53~UT, and show only a few wide-field raster maps for context. The high-cadence program consisted of a three-step raster obtained with the 0.1" slit and a step size of 0.5", observing simultaneously in the \ion{Na}{I} D$_1$ 5896\,\AA~and \ion{Ca}{II} 8542\,\AA~lines (hereafter \ion{Na}{I} D$_1$ and \ion{Ca}{II}), along with several weaker photospheric lines within the same spectral windows. The spectral dispersions and the resolving power were approximately 0.014\,\AA /px and $1.3\times10^5$ for \ion{Na}{I} D$_1$ and 0.018\,\AA /px and $1.5\times10^5$ for \ion{Ca}{II}. 

To improve the signal-to-noise ratio of the polarimetric measurements, the co-aligned Stokes profiles from both ViSP arms were spatially rebinned by a factor of four, to a pixel scale of approximately 0.1" along the slit, resulting in continuum-intensity-normalized uncertainties of $\sigma\approx9\times10^{-4}$ for \ion{Na}{I} D$_1$ and $\sigma\approx1\times10^{-3}$ for \ion{Ca}{II}. These values differ for the context raster. Although the three-step raster was completed in 7.8\,s, instrumental overhead increased the total cadence between successive rasters to 19.6\,s. This corresponds to a Nyquist frequency of $\nu_{\rm Nyq}\approx25.5$\,mHz. The sequence consisted of 110 rasters that span approximately 36 minutes, giving a fundamental frequency spacing of approximately 0.46\,mHz.

\begin{figure}[t]
    \centering
    \includegraphics[width=\linewidth]{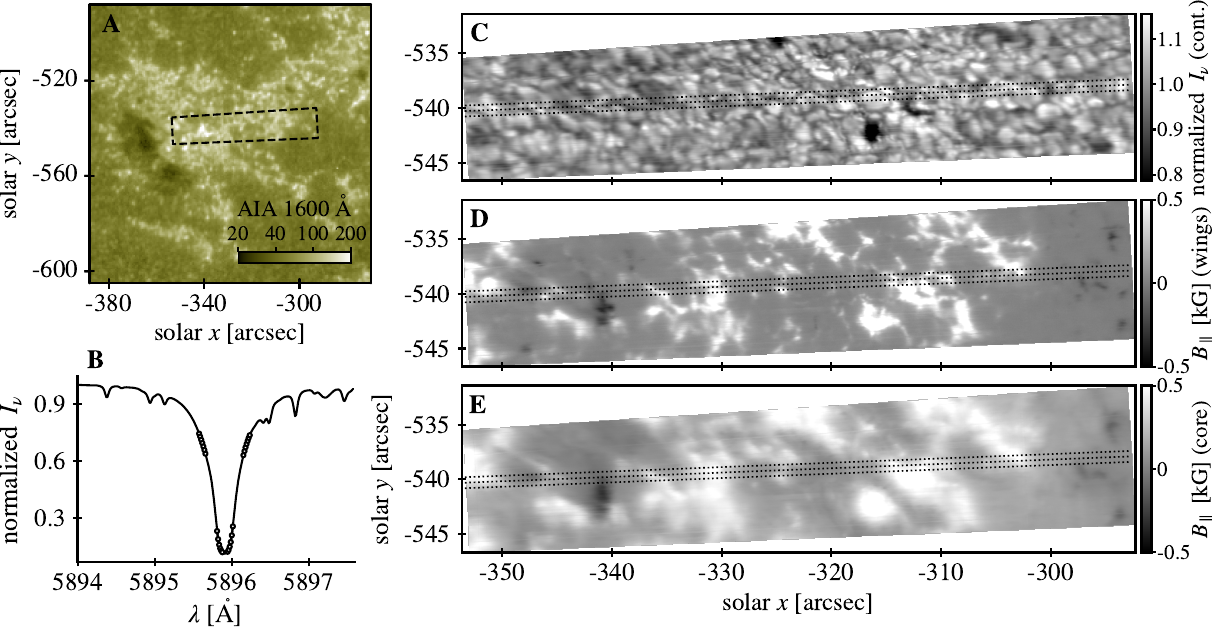}
    \caption{{\bf Overview of the active region plage in 25 July 2025.} (A) SDO/AIA 1600\,\AA~image around 18:44\,UT with square root scaling; the dashed area shows the DKIST/ViSP field of view of the context raster displayed in the right panels. (B) Average continuum-normalized spectrum around the \ion{Na}{I} D$_1$ line; the markers show the wavelength points used for the WFA inversions. (C) \ion{Na}{I} D$_1$ continuum raster map. (D and E) LOS magnetograms at two different height ranges obtained from the wing and core of the \ion{Na}{I} D$_1$ line, as marked in panel (B); the colorbars are capped for display purposes. The three dotted lines show the approximate slit positions in the high-cadence rasters.}
    \label{fig:obs_context}
\end{figure}

Figure~\ref{fig:obs_context} shows an overview of the target as observed by AIA and DKIST. Panel~A displays the AIA 1600~\AA\ image, showing bright plage patches surrounding a trailing sunspot. The dashed box indicates the field of view (FOV) of the context raster, which is largely filled by plage brightenings. The region also contains several pores, i.e., small, dark photospheric magnetic concentrations that lack penumbrae \citep[e.g.,][]{1964suns.book.....B}, which are clearly visible in the ViSP \ion{Na}{I} D$_1$ continuum raster shown in panel~C. This raster began 32~min before the high-cadence program; however, the pores were not sampled by the three-step sequence, as indicated by the dotted lines. The plage region observed by ViSP is dominated by positive-polarity magnetic field (panels~D and E), as inferred from the Stokes~$V$ polarimetry of the \ion{Na}{I} D$_1$ line (panel~B). The magnetic-field estimation is described further in Section~\ref{sec:wfa}. 

\subsection{Line-core Extraction for Velocities and Intensities}
\label{sec:velocities}

Line-core intensity, $\Icore$, and Doppler velocity, $v_{\rm LOS}$, for each pixel of each time step and slit were extracted independently for \ion{Na}{I} D$_1$ and \ion{Ca}{II} Stokes I profiles. For \ion{Na}{I} D$_1$, all profiles contain a single well-defined absorption core (e.g., Fig.~\ref{fig:obs_context}B), so we used an adaptive 11-point quadratic fit centered on the discrete minimum to obtain subpixel estimates of $\lambda_{\rm core}$ and $\Icore$. However, the \ion{Ca}{II} profiles are substantially more complex than the \ion{Na}{I} D$_1$ profiles and frequently exhibit asymmetries, multiple minima, emission shoulders, and partial self-reversals \citep{deLaCruzRodriguez2013, 2020A&A...644A..43P}. Consequently, no single extraction method, including quadratic fitting, center-of-gravity (COG) measurements, or Voigt-profile fitting, performed reliably across the full dataset. Therefore, we adopted an adaptive morphology-aware extraction strategy. Profiles with a clearly defined absorption minimum were measured using automatic adaptive quadratic fits, while highly shifted or strongly distorted profiles were instead manually evaluated using alternative line core measurements selected according to the observed profile morphology; a proper orthogonal decomposition (POD) method was applied to broad or asymmetric profiles, whereas the COG method was used for profiles exhibiting multiple minima or self-reversals, with the adopted solution chosen based on which feature most consistently tracked the line core in the surrounding profiles in space and time.

For the POD method, a local temporal ensemble of profiles at the same spatial position and slit was arranged into a data matrix, and the local mean profile was removed. The resulting matrix, $\mathbf{X}'$, was decomposed using singular value decomposition $\mathbf{X}'=\mathbf{U}\boldsymbol{\Sigma}\mathbf{V}^{\mathsf{T}}$, where the columns of $\mathbf{U}$ describe the temporal coefficients, the diagonal elements of $\boldsymbol{\Sigma}$ rank the relative contribution of each mode and the rows of $\mathbf{V}^{\mathsf{T}}$ contain the corresponding spectral modes. The profile at the selected time step was reconstructed using the three dominant modes:
\begin{equation}
\widehat{I}(\lambda,t)=\overline{I}(\lambda)+\sum_{k=1}^{3}U_{tk}\Sigma_{kk}V_k(\lambda),
\end{equation}
\noindent where $\overline{I}(\lambda)$ is the mean profile of the local ensemble. Retaining only the dominant modes suppresses higher-order noise while preserving the principal line morphology. The reconstructed profile was then smoothed and candidate minima within the restricted \ion{Ca}{II} core window were evaluated according to their depth and proximity to the reference wavelength. The most representative absorption basin was adopted as the line-core position. 

For the COG method, the profile was smoothed within a restricted line-core window and a local continuum level, $I_{\rm cont}$, was estimated from the upper portion of the intensity distribution. The line-core wavelength was then calculated as the intensity-weighted centroid of the absorption depression:
\begin{equation}
\lambda_{\rm COG}=\frac{\displaystyle\sum_i\lambda_iw_i}{\displaystyle\sum_iw_i},\qquad w_i=\max\!\left(I_{\rm cont}-I_i,0\right),
\end{equation}
\noindent where $\lambda_i$ and $I_i$ are the wavelength and intensity at spectral pixel $i$. The non-negative weights emphasize the absorption depression while excluding points above the estimated continuum. The alternative procedures (Supplementary Material Figure~1) were applied only when the default local-minimum extraction did not adequately represent the observed line-core morphology, and the resulting measurements were visually inspected to ensure consistency with neighboring profiles in space and time.

\subsection{Weak-Field Approximation}
\label{sec:wfa}

LOS magnetic-field estimates, $B_{\rm LOS}$ (or $B_{\parallel}$), were derived as a function of time using the weak-field approximation (WFA), which relates Stokes $V$ to the wavelength derivative of Stokes $I$, such that $V(\lambda)\propto B_{\rm LOS}\,(\partial I/\partial\lambda)$ \citep{1973SoPh...31..299L}. The WFA is appropriate because the inferred fields are predominantly weak to moderate ($<1$\,kG), the Zeeman components are not spectrally resolved, the selected Stokes $V$ profiles generally retain the derivative-like morphology assumed by the approximation, and the target is far from the limb \citep[e.g.,][]{2018ApJ...866...89C}.

We used the fast \texttt{spatial\_WFA} code \citep{2020A&A...642A.210M}, which includes two-dimensional spatial regularization. However, because the FOV consisted of only three sparse slit positions, the WFA was applied separately to each slit, effectively regularizing only along the slit direction. The WFA was restricted to wavelength ranges around the line cores similar to those adopted by \citet{2020A&A...642A.210M}. A small wavelength range around the line cores is necessary to mitigate noisy wavelength samples while providing magnetic-field measurements from atmospheric layers comparable to those sampled by the line core Doppler velocities  (Section~\ref{sec:velocities}). Nevertheless, complex or strongly asymmetric Stokes profiles indicative of velocity gradients may depart from the assumptions of the WFA, and the resulting measurements should therefore be regarded as effective line-core field estimates.

A transverse magnetic-field estimate, $B_{\rm TRV}$ (or $B_{\perp}$), was also derived from \ion{Na}{I} D$_1$ and used to estimate the local magnetic-field inclination, $\theta$. The same procedure was not applied to \ion{Ca}{II} because its Stokes signals $Q$ and $U$ were generally too weak to provide reliable transverse-field measurements in the high-cadence rasters. Using a Monte Carlo approach \citep{1992nrfa.book.....P}, in which inversions were repeated for 1000 independent noise realizations, the (statistical) uncertainties in $B_{\rm LOS}$, $\sigma_{B_{\rm LOS}}$, were estimated to be approximately 2\%, or $\sim$5\,G, for \ion{Na}{I} D$_1$ and 10\%, or $\sim$\,20\,G, for \ion{Ca}{II}. This does not include systematic uncertainties due to the assumptions of the WFA \citep[e.g.,][]{2018ApJ...866...89C}.

\subsection{Depth-Stratified Non-LTE Inversions} 
\label{sec:inversions}

We also performed non-local thermodynamic equilibrium (non-LTE) inversions on ViSP data using the Stockholm Inversion Code \citep[\texttt{STiC},][]{2016ApJ...830L..30D,2019A&A...623A..74D}. \texttt{STiC} is a 1.5D radiative-transfer code that assumes plane-parallel geometry and hydrostatic equilibrium, and adjusts a model atmosphere, including temperature, LOS velocity, microturbulent velocity, and magnetic-field parameters as functions of logarithmic continuum optical depth, to match the observed spectra in the least-squares sense. Although waves are inherently dynamic phenomena, simulations indicate that modeling them in a quasi-static manner, by treating each time step independently, provides a reasonable description of the atmospheric response \citep{2021RSPTA.37900182K,2023A&A...670A.133F,daSilvaSantos2024}.

These non-LTE inversions were used primarily to evaluate how velocity determination affects inferred wave flux at different heights. In addition, they provide estimates of the mass densities and geometrical separations between the formation heights of \ion{Na}{I} D$_1$ and \ion{Ca}{II}, which are needed to determine several quantities discussed in Section~\ref{sec:results}, including phase and Alfvén speeds. Although this is a more advanced method than the simpler approaches to determining $v_{\rm LOS}$ and $B_{\rm LOS}$ described in Sections~\ref{sec:velocities} and~\ref{sec:wfa}, it is also far more computationally demanding and more susceptible to noise, which can be problematic for Fourier analysis (Section~\ref{sec:phasescoherence}). Estimates of $v_{\rm LOS}$ and $B_{\rm LOS}$ were based on the line cores for most of the analysis, while non-LTE estimates of $v_{\rm LOS}$ sampled by the wings of both lines were used to calculate the wave flux near the temperature-minimum region (Section~\ref{sec:flux_results}). 

Nevertheless, we find good agreement between the LOS velocities and magnetic fields obtained from different methods (Sections~\ref{sec:nlte_density_height} and \ref{sec:shocks}). In particular, we verified that the magnetic field values are consistent between the different inversion approaches. For example, we find a high degree of linear correlation $r$, between $B_{\rm LOS,~WFA}$ and $B_{\rm LOS,~NLTE}$ throughout the FOV, the highest correlation occurring at $\log \tau=-3.5$ for \ion{Na}{I} D$_1$ ($r=0.98,~p<.001)$ and at $\log \tau=-4.8$ for \ion{Ca}{II} ($r=0.85,~p<.001$). 

\subsection{Power Spectral Density, Cross-spectral Phase, and Coherence} 
\label{sec:phasescoherence}

Power spectral densities (PSDs) were calculated for the $I_{\rm core}$, $v_{\rm LOS}$, and $B_{\rm LOS}$ time series for both lines. Before Fourier analysis, each time series was detrended by subtracting a broad Gaussian-smoothed background with a standard deviation of $\sigma=30$ time samples, equivalent to approximately 9.8~min at the 19.6~s cadence, or an inverse timescale of approximately 1.7~mHz. This procedure suppresses slowly varying long-period trends, but does not impose a sharp frequency cutoff. The detrended time series were also windowed to mitigate spectral leakage and then zero-padded to sample the resulting PSDs on a finer frequency grid. 

Cross-spectral phase differences and magnitude-squared coherence were calculated 
using Fourier-based methods \citep[e.g.,][]{2025NRvMP...5...21J}. Coherence, $C$, was estimated using the well-known Welch method, while representative phase differences were obtained from the complex cross spectrum averaged over the 2.5--5.5\,mHz interval. The analysis was applied to both diagnostics formed within each spectral line, including $\phi(v_{\rm LOS},I_{\rm core})$ and $\phi(B_{\rm LOS},v_{\rm LOS})$, and between the \ion{Na}{I} D$_1$ and \ion{Ca}{II} velocity signals.

 Statistical significance of the magnitude-squared coherence was assessed using a Monte Carlo Fourier-surrogate test based on relative-phase randomization of the individual Welch spectral estimates. Fourier-surrogate methods provide a useful framework for significance testing by preserving selected properties of the observed signals, such as their power spectra, while deliberately destroying the relationship being tested \citep[e.g.,][]{2018PhR...748....1L}. For each spatial pixel, the Welch coherence between two quantities $x$ and $y$ is given by
\begin{equation}
C_{xy}(\nu)=
\frac{|\langle P_{xy}(\nu)\rangle|^2}
{\langle P_{xx}(\nu)\rangle\langle P_{yy}(\nu)\rangle},
\end{equation}
where $P_{xy}$ is the cross-spectrum, $P_{xx}$ and $P_{yy}$ are the corresponding auto-spectra, and the averages are taken over five Welch segments. Surrogate realizations were generated by leaving the observed segment-by-segment auto-spectra unchanged while randomizing the relative phase of each segment cross-spectrum according to $P_{xy,j}^{\rm sur}(\nu) = P_{xy,j}(\nu)e^{i\theta_{j,\nu}},$ and $\theta_{j,\nu}\sim U(-\pi,\pi),$ where $U$ is the uniform distribution. We generated 5,000 such surrogate realizations per pixel. This procedure defines a null hypothesis in which any two quantities ($x$ and $y$) retain their observed segment-by-segment spectral power, while any consistent relative-phase relationship between them is removed. For each surrogate realization, the coherence was averaged over the 2.5--5.5\,mHz frequency range. A 95\% significance threshold was then defined as the 95th percentile of the surrogate null distribution, and only pixels whose observed coherence exceeded this threshold were retained for the phase analysis.

\section{Results}
\label{sec:results}

\subsection{Oscillatory Power and Variance in Plage}
\label{sec:PSDs}

\begin{figure}[t]
\centering
\includegraphics[width=.95\linewidth]{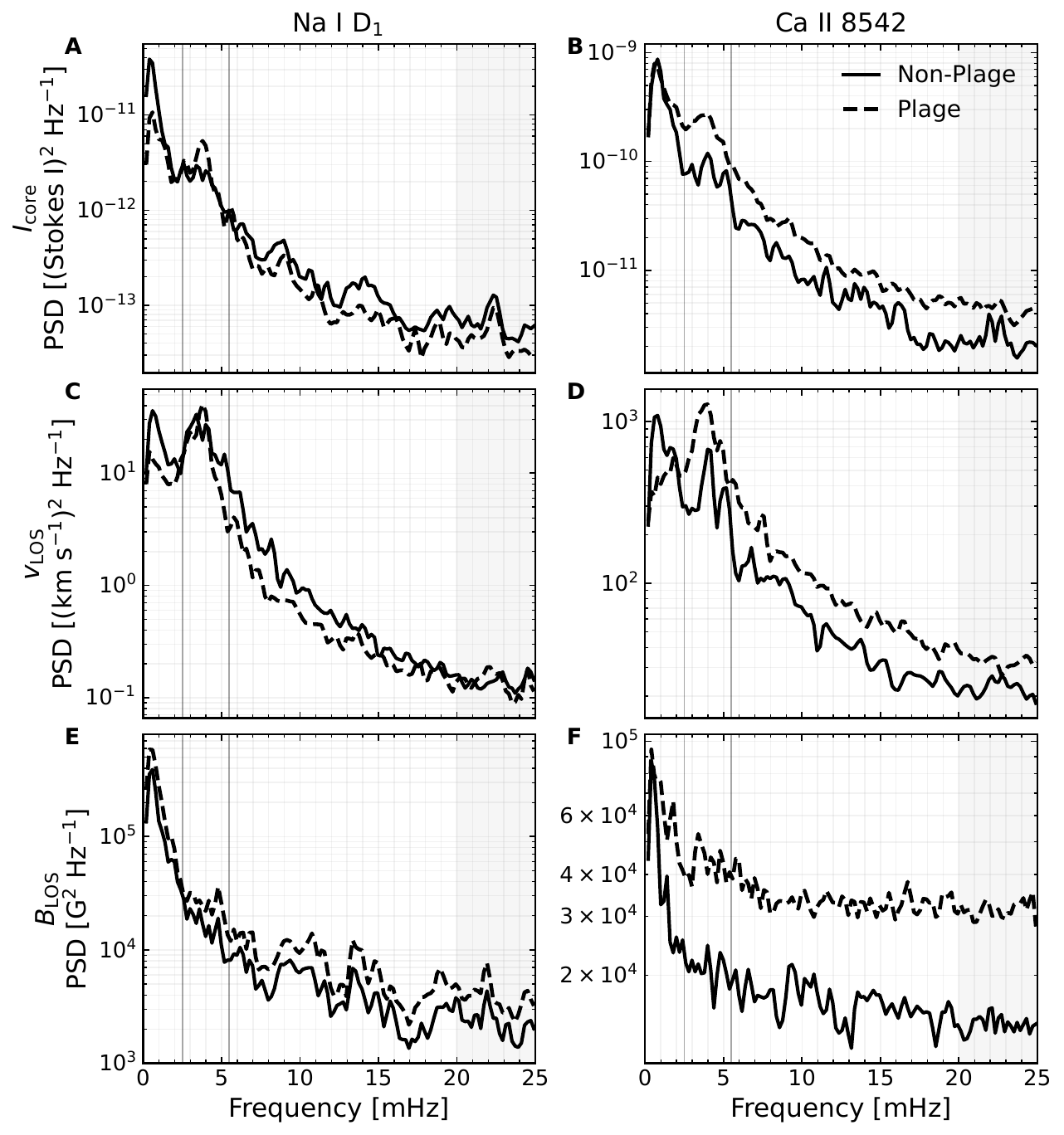}
\caption{\textbf{Average power spectral densities of intensity, Doppler velocity, and LOS magnetic field in plage and surrounding areas.} Panels (A) and (B) show the line-core intensity ($I_{\rm core}$) PSDs for \ion{Na}{I} D$_1$ and \ion{Ca}{II}, respectively; panels (C) and (D) show the corresponding LOS ($v_{\rm LOS}$) PSDs; and panels (E) and (F) show the LOS magnetic-field ($B_{\rm LOS}$) PSDs. Dashed curves denote plage pixels with a signed time-median $B_{\rm LOS}>200$~G in the corresponding spectral line, while solid curves denote non-plage pixels with a signed time-median $B_{\rm LOS}<150$~G. The vertical gray lines mark the adopted $2.5$--$5.5$~mHz frequency band used for the phase and coherence analysis.}
\label{fig:power_spectra}
\end{figure}

Figure~\ref{fig:power_spectra} compares the average PSDs of $I_{\rm core}$, $v_{\rm LOS}$, and $B_{\rm LOS}$ for both lines. Weakly magnetized and plage samples were defined independently for \ion{Ca}{II} and \ion{Na}{I} D$_1$ using the signed time-median $B_{\rm LOS}$, with non-plage pixels having $B_{\rm LOS}<150$~G and plage pixels having $B_{\rm LOS}>200$~G. Both spectral lines exhibit enhanced oscillatory power within the $2.5$--$5.5$\ mHz frequency range in all parameters; however, there are important differences between them, with the enhancement generally stronger in magnetic plage than in the surrounding non-plage regions. 

The exception is the \ion{Na}{I} velocity power, which appears to be suppressed for most frequencies in plage compared to the corresponding \ion{Ca}{II} PSD, showing the opposite behavior (Figure~\ref{fig:power_spectra}C,D). The contrast between plage and non-plage is stronger in \ion{Ca}{II} than in \ion{Na}{I} D$_1$. A possible explanation is that \ion{Ca}{II} samples higher chromospheric layers where waves have larger amplitudes, are more strongly guided by the plage magnetic field, and may have begun to steepen into shocks. This is discussed further in Section~\ref{sec:shocks}. 

The oscillatory signatures are strongest in $v_{\rm LOS}$ and weakest in $B_{\rm LOS}$, although regions of slightly enhanced or approximately flat magnetic power are still present across this frequency range. We note that magnetic oscillations are also detected in the total circular polarization signals $(\int\left|V(\lambda)\right|\,\mathrm{d}\lambda)$, so they are not an artifact of the WFA inversions.  The dominant power enhancements occur generally between 3 and 4 \, mHz for both lines. Motivated by these results, subsequent cross-spectral phase analyses are restricted to the $2.5$--$5.5$\,mHz frequency band. At our frequency resolution,  we find no significant shift of the dominant frequency peaks within the $2.5$--$5.5$\,mHz band toward higher values with height. 

Root-mean-square (RMS) velocity oscillations were obtained taking the square root of the integrated PSDs for both lines in the $2.5$--$5.5$\ mHz band. For plage pixels, we obtained mean values of $\delta v_{\rm LOS,~Na}\approx0.2\rm\,km\,s^{-1}$ and $\delta v_{\rm LOS,~Ca}\approx1.3\rm\,km\,s^{-1}$, corresponding to an apparent increase in velocity amplitude by more than a factor of six between the two atmospheric layers sampled. Similarly, we estimated the RMS magnetic-oscillation amplitude by integrating the $B_{\rm LOS}$ PSDs over the same frequency range and obtained mean values of $\delta B_{\rm LOS,~Na}\approx9$\,G and $\delta B_{\rm LOS,~Ca}\approx8$\,G, with corresponding spatial standard deviations of approximately 4\,G and 3\,G across the FOV. These standard deviations describe the spatial variation among the sampled pixels and are not estimates of the measurement-noise level. 
Similar magnetic RMS values have been observed in the photosphere of pores \citep{Fujimura2009}.

To estimate the contribution of the WFA uncertainties to the measured magnetic power, we propagated the 1,000 Monte-Carlo realizations used to determine the $B_{\rm LOS}$ uncertainties (Section~\ref{sec:wfa}) through the same Fourier analysis. For each noise realization of the $B_{\rm LOS}$ time series at each pixel, we computed the residual relative to the nominal $B_{\rm LOS}$ time series and calculated its PSD. The resulting noise PSD is approximately flat over the $2.5$--$5.5$\,mHz interval, consistent with temporally uncorrelated errors. Direct integration of the Monte-Carlo noise PSD over this band yields an expected RMS noise contribution of $\approx2$\,G for \ion{Na}{I} D$_1$ and $\approx 7$\,G for \ion{Ca}{II}. This is consistent with the analytic expectation for temporally uncorrelated white noise, whose variance is uniformly distributed in frequency up to the Nyquist frequency, $\sigma_{2.5-5.5}=\sigma_{B_{\rm LOS}}\sqrt{\Delta \nu/\nu_{\rm Nyq}}$, where $\sigma_{B_{\rm LOS}}$ are the WFA uncertainties (Section~\ref{sec:wfa}). Thus, the measured \ion{Na}{I} D$_1$ magnetic RMS amplitude lies well above the expected contribution from the WFA measurement uncertainties, whereas the \ion{Ca}{II} result is comparable to the corresponding noise level and should therefore be interpreted with caution. A formal quadrature subtraction yields an intrinsic \ion{Ca}{II} RMS amplitude of $\sim4$\,G.

\subsection{NLTE Constraints on Density and Na--Ca Formation-Height Separation}
\label{sec:nlte_density_height}

\begin{figure}[t]
    \centering
    \includegraphics[width=\linewidth]{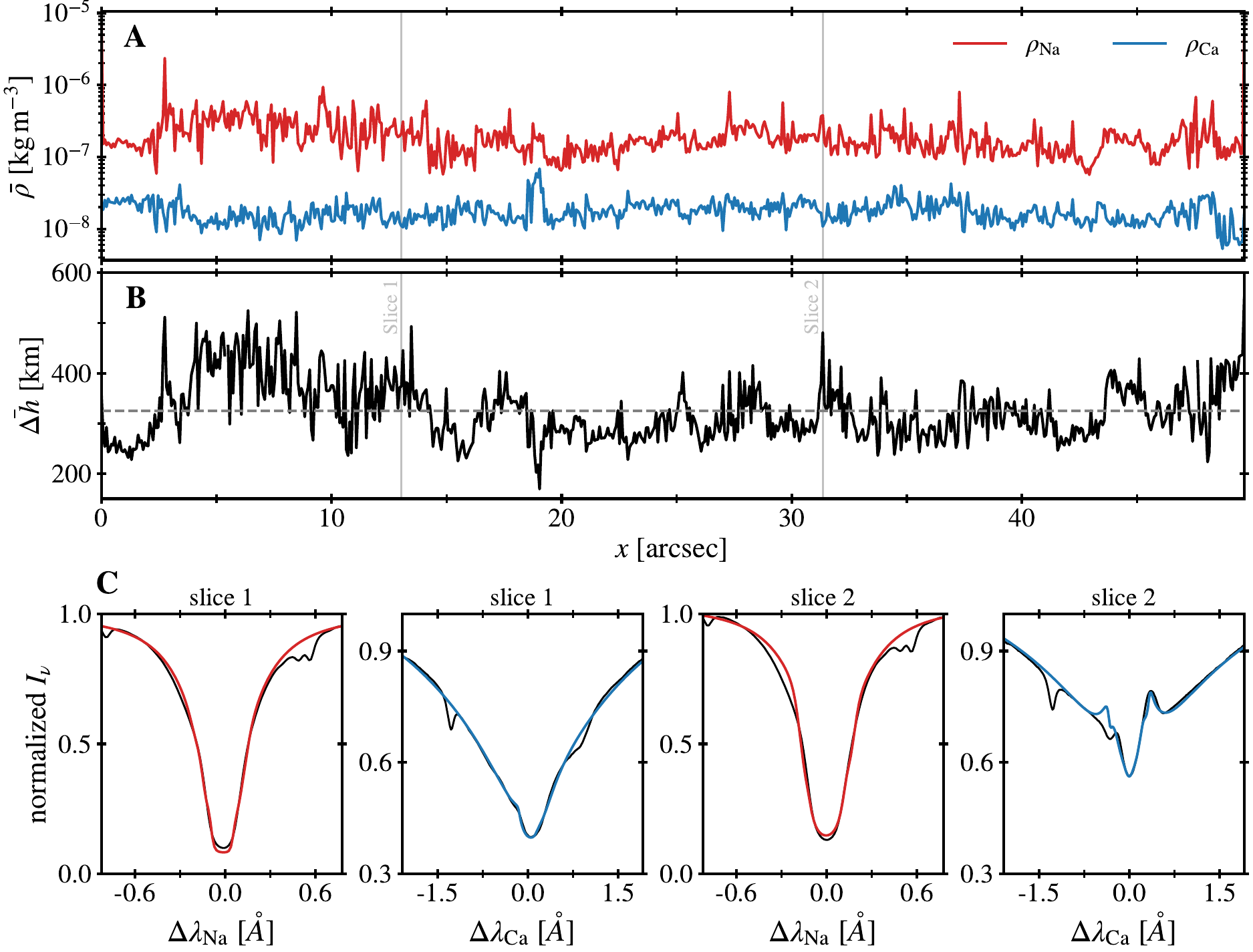}
    \caption{{\bf Average mass densities and \ion{Na}{I} D$_1$--\ion{Ca}{II} separation heights from non-LTE inversion models.} (A): Mass densities along the slit direction at the $\tau=1$ layer of the \ion{Na}{I} D$_1$ and \ion{Ca}{II} line cores, averaged across the three slits. (B): Height separation between the two line cores averaged across the three slits; the dashed line shows the average value. (C): Example observed average spectra (black) and best fits (colored) at two different locations along the slit marked in the top panels.}
    \label{fig:densities}
\end{figure}

Figure~\ref{fig:densities} shows the mass densities evaluated at the optical-depth-unity surfaces of the two line cores and the corresponding geometrical separation between these surfaces along the slit direction, as inferred from the non-LTE inversions described in Section~\ref{sec:inversions}. The $\tau_{\rm core}=1$ layer was chosen because the three main diagnostic quantities ($I_{\rm core}, v_{\rm LOS}, B_{\rm LOS}$) were extracted at or near the line cores. We note that the mass density is not a free parameter in the inversion but is set by the equation of state and the constraint of hydrostatic-equilibrium \citep{2019A&A...623A..74D}. In addition, the assumption of hydrostatic equilibrium neglects the influence of magnetic forces on the atmospheric pressure scale height. These approximations are most appropriate for estimating the mean background stratification and may not hold instantaneously during strongly nonlinear wave propagation. 

The results indicate that the characteristic density ordering remains relatively stable: \ion{Na}{I} D$_1$ consistently samples a denser and lower atmospheric region than \ion{Ca}{II} with average values of $\rho\sim$\,$2.2(\pm0.6)\times10^{-7}$ and $\rho\sim$\,$1.8(\pm2.0)\times10^{-8}$~kg\,m$^{-3}$ (Figure~\ref{fig:densities}A). For comparison, using the same atomic data as for inversions, the semi-empirical plage model by \citet{2009ApJ...707..482F} yields $\rho^{\rm F09}_{\rm Na}\sim7.4\times10^{-7}$ and $\rho^{\rm F09}_{\rm Ca}\sim1.2\times10^{-8}$~kg\,m$^{-3}$. The latter value agrees with that inferred from our inversions, whereas the former appears too high, and it would not reproduce our observations. Using the F09 values would increase the wave fluxes between the Na--Ca layers by a factor of three (Section~\ref{sec:flux_results}).

The average geometric separation (standard deviation) between the line-core sampling heights of \ion{Na}{I} D$_1$ and \ion{Ca}{II} is $\Delta h=325 (\pm60)$~km in our inversion models (Figure~\ref{fig:densities}B). Our $\Delta h$ is approximately half the value reported by \citet{2020A&A...642A.210M}; adopting their larger height separation would proportionally increase the inferred wave phase speeds (Section~\ref{sec:phasespeeds}). However, they calculated $\Delta h$ from the difference between the centers of gravity of the response functions over a range of wavelengths near the line cores, rather than from the strict $\tau_{\rm core}=1$ layers of the two lines.

Although formation-height variations associated with changes in density and magnetic topology might reasonably be expected \citep[e.g.,][]{2023ApJ...945..154M}, we did not find a clear correlation between the spatial profiles of density or height separation and the underlying LOS magnetic-field strength. As with the WFA, we note that the transverse field inferred from the non-LTE inversions is generally unreliable at chromospheric depths. Longer exposures or a greater number of modulation cycles would be required, at the expense of cadence. Using $|B_{\rm LOS}|$ as a proxy for the total field strength, both line cores generally form above the plasma $\beta=1$ layer in our models. Because our non-LTE models generally provide reasonable fits to the observed Stokes I spectra (Fig.~\ref{fig:densities}C), we use the inferred average values of $\rho$ and $\Delta h$ as a more physically consistent basis for the wave analysis than values adopted from the literature, while retaining the more robust velocities independently derived using the adaptive method to mitigate the effects of inversion noise in the phase analysis.

\subsection{\ion{Na}{I} D$_1$ -- \ion{Ca}{II}  Velocity Phase Differences, Coherence, and Spatial Magnetic Structure}
\label{sec:phasespeeds}

\begin{figure}[ht]
    \centering
    \includegraphics[width=0.81\linewidth]{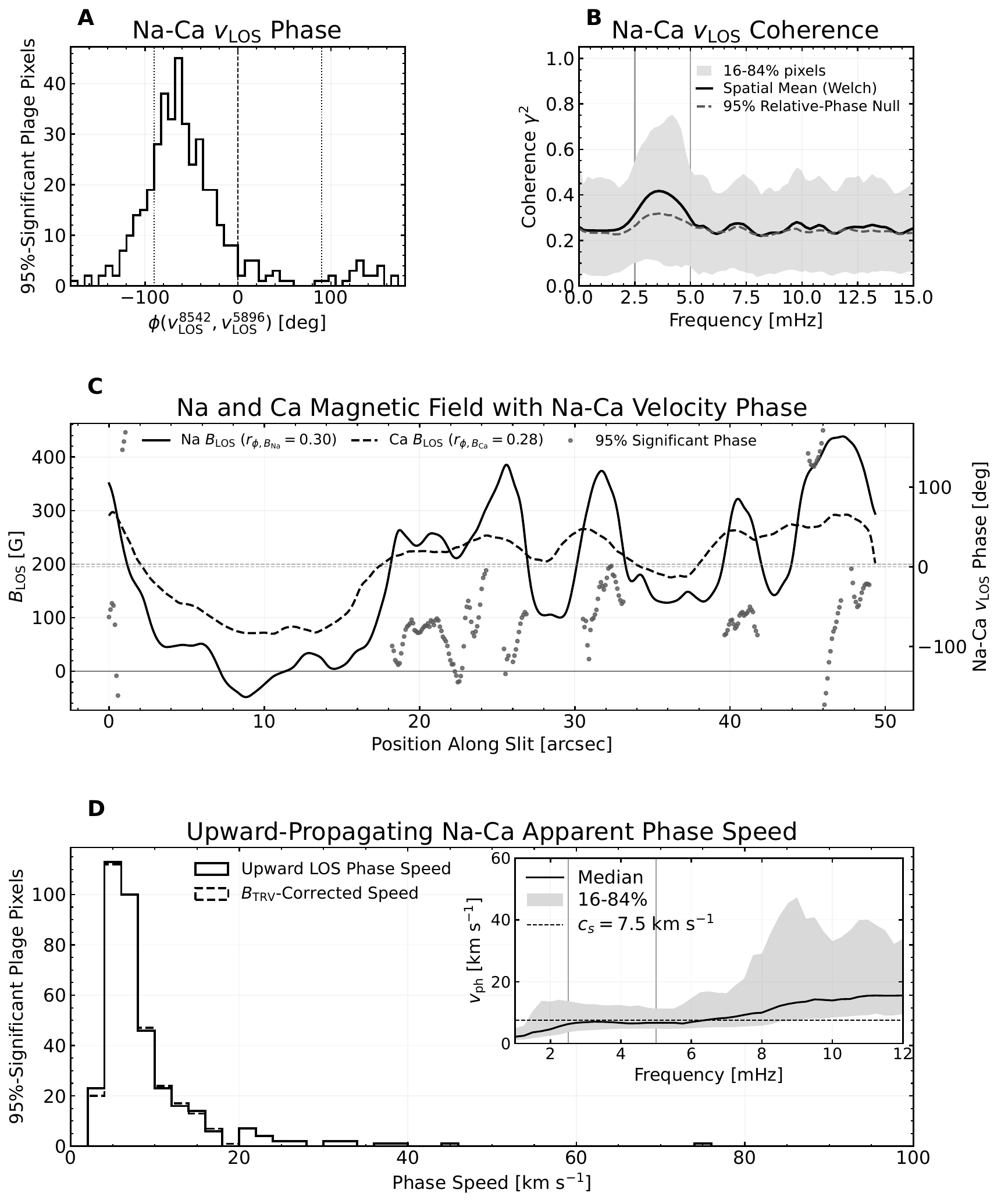}
    \caption{
    \textbf{Cross-spectral Na--Ca phase relationships.}
    (A) Histogram of \ion{Na}{I} D$_1$--\ion{Ca}{II} $v_{\rm LOS}$ phase differences for plage pixels whose band-averaged coherence over 2.5--5.5\,mHz exceeds the 95\% significance threshold. 
    (B) Welch coherence spectrum as a function of frequency. The gray region shows the 16th--84th percentile range across pixels, the black curve shows the spatial-mean coherence, and the dashed curve shows the 95th percentile of the relative-phase randomized null distribution. The vertical shaded region indicates the 2.5--5.5\,mHz analysis band.
    (C) Spatial variation of the \ion{Na}{I} D$_1$ and \ion{Ca}{II} $B_{\rm LOS}$ along the central slit, together with the Na--Ca velocity phase difference, $\phi(v_{\rm LOS}^{\rm Ca},v_{\rm LOS}^{\rm Na})$, for pixels satisfying the 95\% significance criterion.
    (D) Phase speed for the significant plage pixels within the 2.5--5.5\,mHz band; the inset shows the phase speed as a function of frequency.
    }
    \label{fig:NaCa_phase_coherence}
\end{figure}

To characterize the frequency-dependent coupling between the \ion{Na}{I} D$_1$ and \ion{Ca}{II} velocity signals, we performed a series of complementary phase, coherence, magnetic-field, and apparent phase-speed comparisons. We first computed the cross-spectrum between the two LOS time series for each plage pixel (Figure~\ref{fig:NaCa_phase_coherence}A) and used the corresponding Welch coherence to identify statistically-significant inter-line phase measurements (Figure~\ref{fig:NaCa_phase_coherence}B) at the 95\% level. 

The Na--Ca velocity phase difference was defined as $\phi_{\rm Na-Ca}=\phi(v_{\rm LOS}^{\rm Ca})-\phi(v_{\rm LOS}^{\rm Na})$, such that negative values indicate that the upper \ion{Ca}{II} signal lags the lower \ion{Na}{I} D$_1$ signal. The phase histogram is strongly concentrated at negative phase angles, indicating that the \ion{Ca}{II} velocity signals systematically lag the \ion{Na}{I} D$_1$ signals (Fig.~\ref{fig:NaCa_phase_coherence}A). The spatially averaged Na--Ca \(v_{\rm LOS}\) coherence shows a clear enhancement over 2.5--5.5\,mHz, reaching a moderate peak of \(C\sim0.42\), but lying well above the 95\% relative-phase-randomized null level, indicating a statistically significant phase relationship between the two velocity signals in this frequency range.

The phase lag in $v_{\rm LOS}$ between the Na--Ca layers was converted to a travel time using $\tau=|\Delta\phi|P/360^{\circ}$, with $P=1/\nu_{\rm eff}$ determined from the coherence-weighted cross-power frequency. The mean travel time is $49\pm17$~s in significant plage pixels ($N=126$) and $44\pm21$~s in significant non-plage pixels ($N=34$). Including all upward-propagating non-plage pixels without a coherence threshold gives $66\pm37$~s ($N=294$). We interpret the small secondary population near $+140^\circ$ as a downward-propagating component.

To examine how the propagation properties vary across magnetic structures, Fig.~\ref{fig:NaCa_phase_coherence}C compares the \ion{Na}{I} D$_1$ and \ion{Ca}{II} $B_{\rm LOS}$ measured along the central slit with the corresponding Na--Ca velocity phase difference.  Visually, the Na--Ca v$_{LOS}$ phase is associated with stronger magnetic concentrations, indicating that the plage magnetic elements act as efficient waveguides for the observed Na--Ca propagation. We also tested whether the spatial gradients of $\phi_{\rm Na-Ca}$ and $|B_{\rm LOS}|$ were statistically correlated across the slit and found that the phase and magnetic-field profiles show a weak-to-moderate positive correlation. However, there is appreciable spatial co-variability of the measured phase lag and LOS magnetic field across some localized regions. For example, between approximately $x=22$--$28$~arcsec, the phase-lag and magnetic-field gradients were positively correlated, reaching $r=0.92$ for \ion{Na}{I} D$_1$ and $r=0.79$ for \ion{Ca}{II}. These correlation values suggest that some local phase variations and wave travel times may be associated with the strongest magnetic-field gradients, likely as a function of the local inclination angle. 

The phase speed was estimated as $v_{\rm ph}=(2\pi \nu\,\Delta h)/|\Delta\phi|$, where $\Delta h$ is the assumed formation-height separation and $|\Delta\phi|$ is the measured Na--Ca velocity phase difference. Figure~\ref{fig:NaCa_phase_coherence}D shows that the dominant apparent phase-speed population lies near $7.5~\mathrm{km\,s^{-1}}$ within the $2.5$--$5.5$\,mHz band, comparable to the expected chromospheric sound speed within the uncertainties. Phase-speed populations above and below $20~\mathrm{km\,s^{-1}}$ occur within the same 2.5--5.5\,mHz interval (Supplementary Material Figure~2). 

The local Alfv\'en speed was estimated as $v_A=|B_{\rm LOS}|/\sqrt{\mu_0\rho}$ using the time-median LOS magnetic field. The mass densities at the two formation depths were taken from the non-LTE inversions of the same dataset (Section~\ref{sec:inversions}). For statistically significant pixels displaying upward propagation, the estimated median lower-limit Alfv\'en speeds are approximately $35$~km\,s$^{-1}$ at the \ion{Na}{I} D$_1$ height and $110\pm10$~km\,s$^{-1}$ at the \ion{Ca}{II} height, with corresponding standard deviation values of approximately 14 and 10~km\,s $^{-1}$, respectively. These are estimates of the lower-limit because they are calculated using $B_{\rm LOS}$ rather than the total magnetic-field strength. The conservative two-layer threshold is therefore set by the smaller \ion{Na}{I} D$_1$ layer value at each pixel. Approximately $98\%$ of the selected pixels have apparent phase speeds below this lower-limit Alfv\'en speed. The predominance of sub-Alfv\'enic apparent phase speeds is consistent with the dominant population being associated with slow magnetoacoustic propagation.

\subsection{Slow Compressive and Sausage-Like Internal Phase Signatures}
\label{sec:phases}

\begin{figure}[t]
\centering
\includegraphics[width=\linewidth]{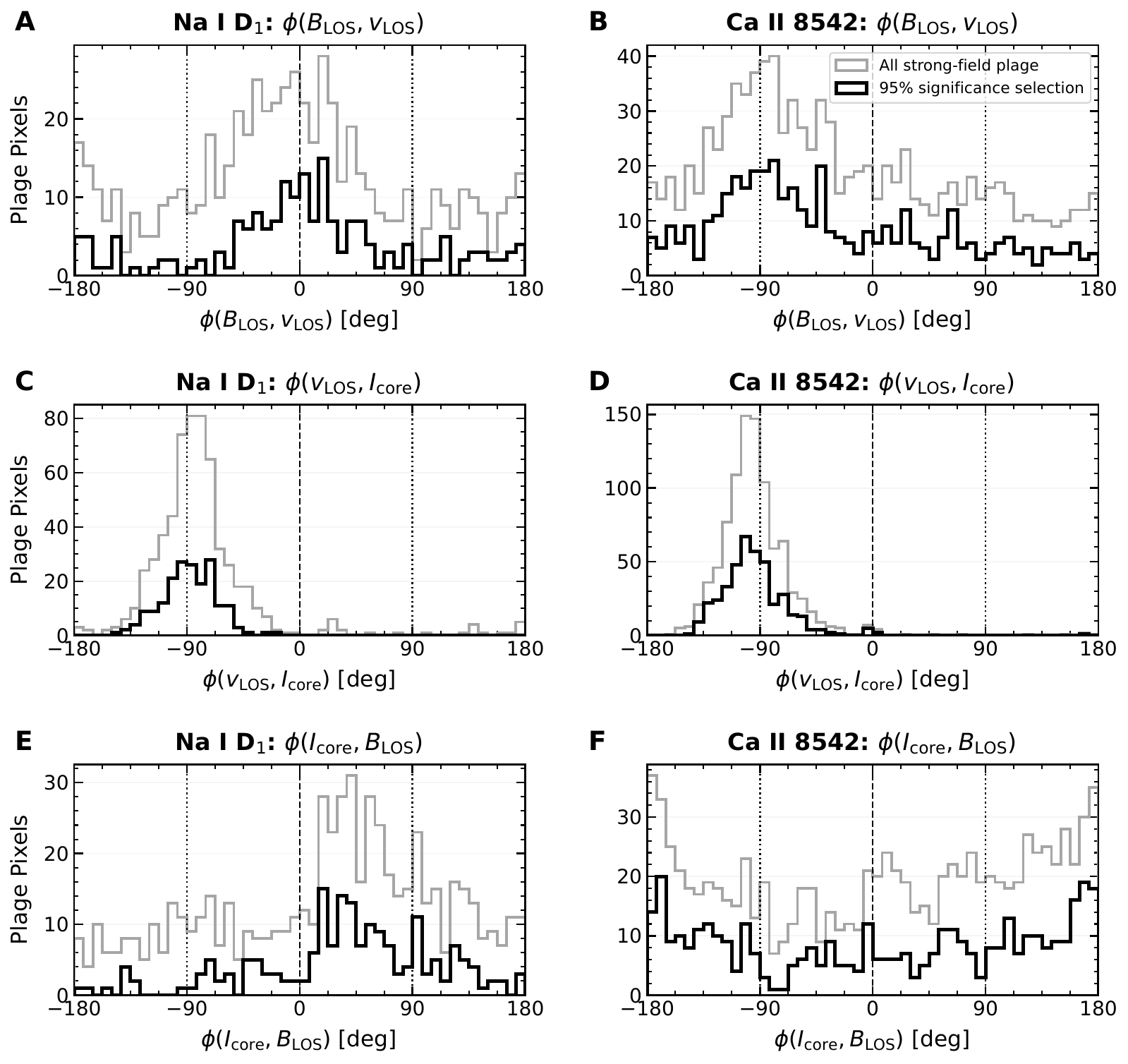}
\caption{
\textbf{Internal phase relationships for plage pixels in the 2.5--5.5\,mHz band.}
Panels (A)--(F) show the internal phase distributions between LOS magnetic field, Doppler velocity, and line-core intensity for \ion{Na}{I} D$_1$ (left column) and \ion{Ca}{II} 8542\,\AA\ (right column): (A,B) $\phi(B_{\rm LOS},v_{\rm LOS})$, (C,D) $\phi(v_{\rm LOS},I_{\rm core})$, and (E,F) $\phi(I_{\rm core},B_{\rm LOS})$. Gray histograms show all valid strong-field plage pixels with a time-median $|B_{\rm LOS}|>200$~G in the corresponding spectral line, while black histograms show the subset satisfying the pixel-specific 95\% coherence confidence level in all three internal observable pairs for that line. The vertical dashed line denotes $0^\circ$, while the dotted lines denote $\pm90^\circ$.}
\label{fig:internal_phase}
\end{figure}

Figure~\ref{fig:internal_phase} shows the internal phase distributions for plage pixels within the 2.5--5.5\,mHz band. For these internal phase diagnostics, coherence significance is evaluated separately for each pixel and observable pair using the same Fourier random-phase surrogates method applied to the $v_{\rm LOS}-v_{\rm LOS}$ relationship (Section~\ref{sec:phasespeeds}). Pixels are retained when their 2.5--5.5\,mHz band-averaged Welch coherence exceeds the corresponding pixel-specific 95th-percentile surrogate threshold in all three internal observable pairs for that spectral line. Overall, the selected phase distributions show preferred relationships among the magnetic, velocity, and intensity perturbations within the dominant oscillation band. Although the significance selection criterion reduces the number of pixels, the main features of the distributions are essentially preserved. This indicates that the inferred phase relationships are not introduced by the coherence selection itself.

 \citet{Moreels2013} considered linear MHD waves in a straight, uniform cylindrical magnetic flux tube, without invoking the thin-tube approximation. However, the effect of density stratification is not included in the model. For example, for a slow standing sausage mode, their model predicts LOS phase relationships $\phi_B-\phi_v=\pm90^\circ$, $\phi_v-\phi_I=\pm90^\circ$, and $\phi_I-\phi_B=180^\circ$, whereas a slow propagating sausage mode gives $(\phi_B-\phi_v,\phi_v-\phi_I,\phi_I-\phi_B)
=(180^\circ,0,180^\circ)$. Fast modes would also show distinct relationships \citep[see Table 1 in][]{Moreels2013}. We benchmark our measurements against these predictions.

Although noisier, the \ion{Ca}{II} phase relationships (Figure ~\ref{fig:internal_phase}B,D,F) provide the clearest indication of the distinct height-dependent slow standing or partially standing surface sausage-like wave phase behavior in magnetic flux tubes \citep{Moreels2013}. Both the $B_{\rm LOS}$ -- $v_{\rm LOS}$ and $v_{\rm LOS}$ -- $I_{\rm core}$ distributions are concentrated near $-90^\circ$ (Figure~\ref{fig:internal_phase}B,D), suggesting that the line responds to a common underlying wave process and is consistent with oscillations exhibiting a compressive component along the LOS. The $I_{\rm core}$ -- $B_{\rm LOS}$ phase distribution is preferentially enhanced near the circular phase boundaries at $\pm180^\circ$, which represents an anti-phase relationship (Figure~\ref{fig:internal_phase}F). Therefore, the \ion{Ca}{II} phase distributions are internally consistent and satisfy the phase-closure relation. 

The distributions in \ion{Na}{I} D$_1$ show distinct magnetic phase offsets that are superposed on the common intensity--velocity coupling (Figure~\ref{fig:internal_phase}A,C,E). The magnetic phase distributions in \ion{Na}{I} D$_1$ qualitatively overlap with the fast propagating surface sausage-mode wave solution \citep{Moreels2013}. In particular, the histogram for \ion{Na}{I} D$_1$ $B_{\rm LOS}$ -- $v_{\rm LOS}$ contains a broad central population centered at approximately $0^\circ$, together with weaker contributions centered around $\pm180^\circ$ (Figure~\ref{fig:internal_phase}A). The corresponding distribution $I_{\rm core}$-- $B_{\rm LOS}$ appears to be dominated by a bi-modal phase population offset above $0^\circ$ around $35^\circ$, with an additional weaker structure near $-90^\circ$ (Figure~\ref{fig:internal_phase}E). One possible explanation is that the \ion{Na}{I} D$_1$ core samples layers much closer to the $\beta\sim1$ region than \ion{Ca}{II} 8542, where fast--slow coupling can produce mixed signatures \citep[e.g.,][]{SchunkerCally2006}.

\subsection{\ion{Na}{I} D$_1$ and \ion{Ca}{II} Shock Signatures}
\label{sec:shocks}

\begin{figure}[t]
    \centering
    \includegraphics[width=\textwidth]{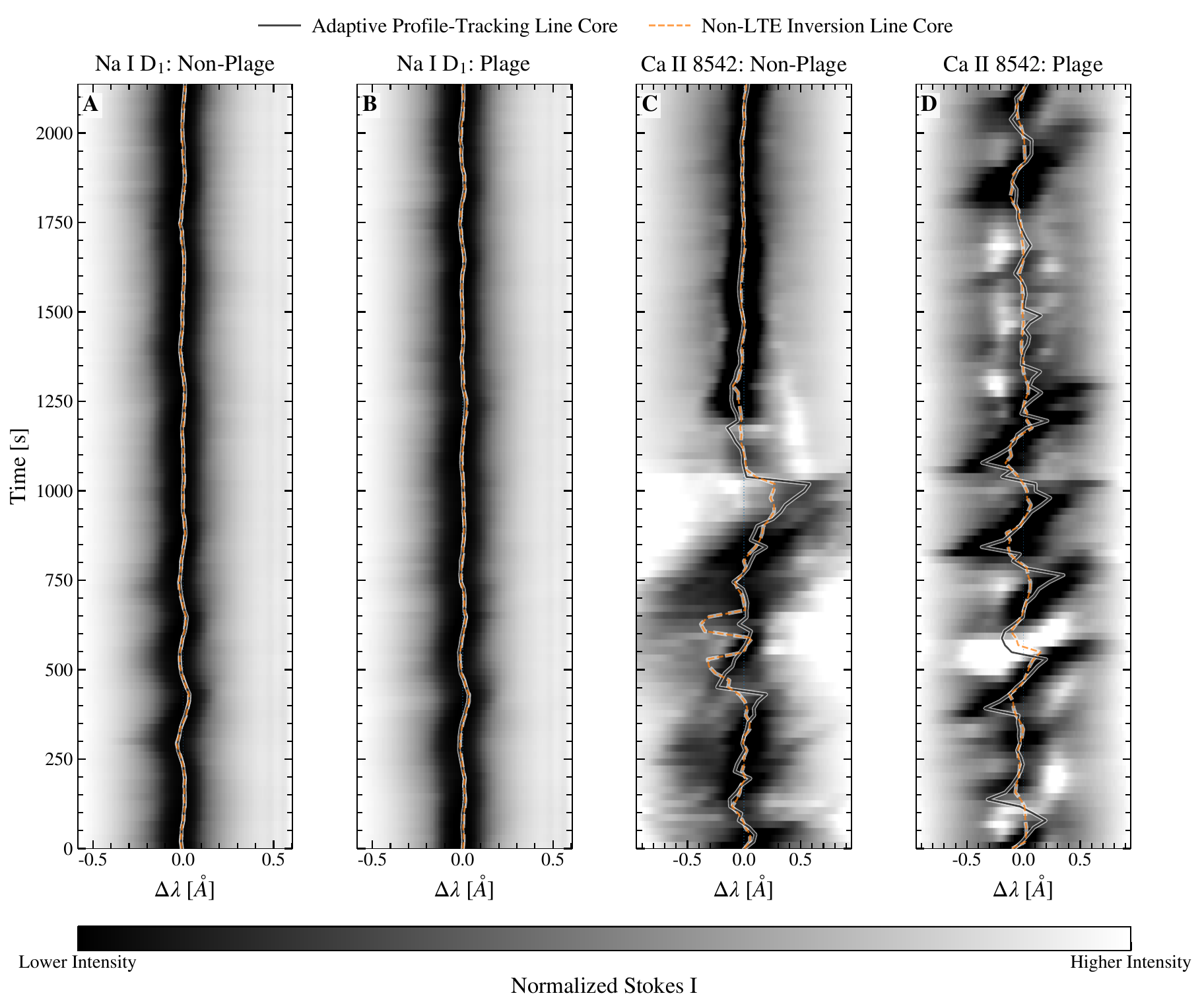}
    \caption{
    \textbf{Representative wavelength--time diagrams for \ion{Na}{I} D$_1$ and \ion{Ca}{II}.}
    Panels (A) and (B) show non-plage and plage \ion{Na}{I} D$_1$ profiles, respectively, while panels (C) and (D) show the corresponding non-plage and plage \ion{Ca}{II} profiles. The solid curves trace the line-core positions obtained with the adaptive profile-tracking extraction, while the orange dashed curves show the line-core positions inferred from the non-LTE inversion velocities. 
    }
    \label{fig:lambda_time_shocks}
\end{figure}

To assess whether the observed chromospheric velocity fluctuations are associated with shock-like spectral evolution, we examined wavelength--time diagrams constructed from the Stokes $I$ profiles of \ion{Na}{I} D$_1$ and \ion{Ca}{II} in both non-plage and plage regions. Figure~\ref{fig:lambda_time_shocks} shows selected representative examples from both magnetic environments, although the same trends are seen throughout the rest of the FOV. Each panel displays the temporal evolution of the line profile as a function of wavelength offset from the reference line center. The adaptive profile-tracking line cores are over-plotted together with the corresponding line core positions inferred from the non-LTE inversions for comparison. 

The \ion{Na}{I} D$_1$ profiles exhibit comparatively smooth and slowly varying wavelength displacements in non-plage and plage regions, with no abrupt, sawtooth-like excursions characteristic of shocks \citep[e.g.,][]{1997ApJ...481..500C}. The adaptive and inversion-derived line-core tracks remain closely aligned throughout the time sequence, supporting the reliability of the extracted \ion{Na}{I} D$_1$ Doppler evolution. 

In contrast, the \ion{Ca}{II} wavelength--time diagrams show rapid line-core displacements, strong profile asymmetries, and abrupt transitions between redshifted and blueshifted states, particularly in plage regions, where these signatures occur more frequently in an intermittent way. The adaptive profile-tracking and non-LTE inversion curves reproduce the same large-scale excursions, although short-lived differences arise during the most asymmetric or rapidly evolving profiles. 

The difference between the two diagnostics is consistent with their distinct atmospheric sensitivities, with \ion{Na}{I} D$_1$ sampling a comparatively lower and more weakly shocked layer than the higher atmospheric layers probed by \ion{Ca}{II}, where the density is approximately one order of magnitude lower (Section~\ref{sec:nlte_density_height}). The agreement between the independently derived line-cores indicates that these excursions are not artifacts of the profile-tracking procedure. This nonlinear evolution may contribute to the scatter in the linear phase relationships presented in Section~\ref{sec:phases} for \ion{Ca}{II}, while the smoother evolution of \ion{Na}{I} D$_1$ is less likely to be affected. 

\subsection{Variation of Wave-Flux with Height}
\label{sec:flux_results}

We calculated the kinetic component of the wave energy flux ($\rho\,\delta v^2_{\rm LOS}\,v_{
\rm gr}$) at different heights as follows:
\begin{equation}
    F=\rho\,\int \frac{[P_{v}(\nu)-P_{\rm noise}]\,v_{\rm gr}(\nu)}{\mathcal{T}(\nu)}\,d\nu
\end{equation}
\noindent where $P_v(\nu)$ is the velocity PSD as a function of frequency, $P_{\rm noise}$ is an empirical estimate of the approximately frequency-independent high-frequency noise floor, taken as the median velocity power over $20$--$25$~mHz where the PSD flattens, and $v_{\rm gr}(\nu)=c_s\sqrt{1-(\nu_{\rm ac}/\nu)^2}$ is the group speed. The frequency-dependent transfer function, $\mathcal{T}({\nu})$, was assumed to be unity \citep[e.g.,][]{2010ApJ...723L.134B,Sobotka2016,Abbasvand2020b}, since we do not know its behavior in plage regions. Wave fluxes were integrated over the broader $2.5$--$20$\ mHz spectral interval. Frequencies higher than our 2.5--5.5\,mHz range were included in this calculation, as they may be relevant to chromospheric heating \citep[e.g.,][and references therein]{Srivastava2021}. 

To estimate the plausible cutoff range, we used the inversion-derived thermodynamic variation in the lower layer, which gives vertical acoustic cutoffs of approximately $4.72$--$5.26$~mHz. Applying the $\cos\theta$ inclination correction \citep[e.g.,][]{SchunkerCally2006}, where $\theta$ is measured relative to the solar vertical after correcting for the $\sim41^{\circ}$ heliocentric viewing angle, yields effective cutoffs of approximately $3.27$--$5.11$~mHz. We adopted $\nu_{\rm ac}=3.5$~mHz as a common effective cutoff for the wave-flux calculation.  This conservative choice maximizes the included low-frequency wave power. We retained only pixels with Na--Ca $v_{\rm LOS}$ phases corresponding to upward propagation, and tests using pixel-dependent inclination-corrected cutoffs reduce the absolute fluxes but do not alter our conclusions.

To estimate wave energy fluxes through the lower chromosphere  for comparison with the literature, we extracted mass densities at $\log \tau=-4$ (referred to as {\it low}), which is the average depth of the temperature minimum at the base of the chromosphere, and at $\log \tau=-5$ (referred to as {\it high}), the average formation height of the \ion{Ca}{II} line from the non-LTE inversions (Section~\ref{sec:inversions}). The temperature-minimum layer is constrained by the intensities in the wings of both lines and therefore cannot be probed using our line-centroid method (Section~\ref{sec:velocities}). The average ($\pm$ standard deviation) density at that depth is $\rho_{\rm low}\approx1.6(\pm0.6)\times10^{-6}$~kg\,m$^{-3}$.  

\begin{figure}[t]
\centering
\includegraphics[width=\textwidth]{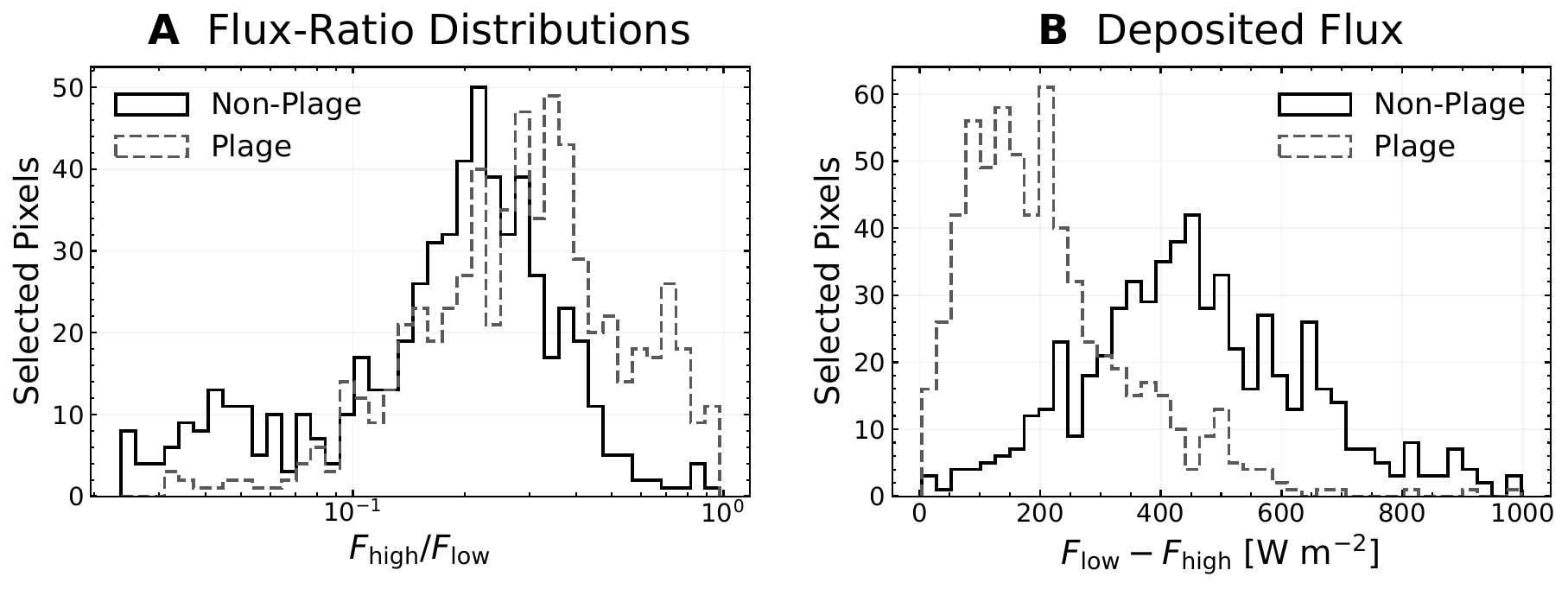}
\caption{{\bf Wave-flux ratios and deposited fluxes derived from the non-LTE inversion velocities for the upward-propagating non-plage and plage populations.} Fluxes were evaluated using non-LTE inversion-derived velocities and densities at $\log\tau=-4$ and $-5$, a fixed effective cutoff of 3.5~mHz, and noise-subtracted power over 2.5--20~mHz. Only pixels with Na--Ca $v_{\rm LOS}$ phases corresponding to upward propagation were retained. No coherence threshold was imposed. (A) Distribution of the upper-to-lower flux ratio, $F_{\rm high}/F_{\rm low}$. (B) Distribution of the deposited flux, $F_{\rm low}-F_{\rm high}$. Solid black curves denote non-plage pixels, while dashed gray curves denote plage pixels.
}
\label{fig:flux}
\end{figure}

The flux ratio and the deposited flux were calculated as $F_{\rm high}/F_{\rm low}$ and $F_{\rm low}-F_{\rm high}$, respectively. Figure~\ref{fig:flux} shows histograms of the (acoustic) wave-flux ratios and deposited wave-energy fluxes derived from non-LTE inversions. The plage population exhibits larger $F_{\rm high}/F_{\rm low}$ ratios and smaller deposited fluxes because both $F_{\rm low}$ and $F_{\rm high}$ are lower than in non-plage, with $F_{\rm low}$ decreasing much more strongly, leaving a larger fraction of the available upward wave flux at the upper height. We find relatively low deposited acoustic fluxes with median values $\sim$\,180\,$\rm W\,m^{-2}$ in plage and $\sim$\,450\,$\rm W\,m^{-2}$ in non-plage,  with an estimated uncertainty of 40\%, arising primarily from the uncertainty in the mass density.

Higher deposited wave fluxes have been reported in plage regions observed closer to disk center, although the height ranges considered differ somewhat among studies \citep[e.g.,][]{Sobotka2016,2019ApJ...871..155R,Abbasvand2020b,Morosin2022,daSilvaSantos2024}. We note that if we calculated the flux deposited between the narrower optical depth range between the Na--Ca $\tau_{\rm core}=1$ layers using $\rho_{\rm Na}\approx2.2(\pm0.6)\times10^{-7}$~kg\,m $^{-3}$ (Section~\ref{sec:nlte_density_height}), the result would instead indicate an apparent amplification of the wave flux, because the larger \ion{Ca}{II} velocity RMS more than compensates for the approximately tenfold decrease in density between the two layers. 

Non-LTE inversion derived velocities may  underestimate the velocity power \citep{daSilvaSantos2024}, while simulations suggest that the transfer function is slightly smaller than unity \citep{2023ApJ...945..154M}, both of which would increase our estimated wave fluxes. Adopting a fixed cutoff of 3.5~mHz may overestimate the propagating contribution where the local cutoff is higher. We therefore regard the resulting fluxes as approximate estimates. Nevertheless, the energy requirements of ARs are canonically $\gtrsim$10\,${\rm kW\,m^{-2}}$ on average in the low chromosphere \citep{Withbroe1977}, with high-resolution observations suggesting even higher radiative losses with substantial spatial structure \citep[up to $\sim$20\,${\rm kW\,m^{-2}}$;][]{Morosin2022}. These values exceed all reported wave-flux measurements in AR plage.

\section{Discussion}

The internal $v_{\rm LOS}$--$I_{\rm core}$ phase distributions in both \ion{Na}{I} D$_1$ and \ion{Ca}{II} (Figure \ref{fig:internal_phase} C,D) show coherent compressive oscillations in the dominant $2.5$--$5.5$\ mHz band, corresponding to periods of approximately 3--7 minutes. Periods between 3--7 minutes are consistent with slow magnetoacoustic waves that have been observed to propagate from the photosphere to the transition region in plage \citep[e.g.,][]{2003ApJ...595L..63D, Kayshap2020, 2025MNRAS.543.3791C}. To assess whether the observed $v_{\rm LOS}$--$I_{\rm core}$ phase shows a delayed radiative response of the line-core intensity, we examined the phase between the inversion-derived $v_{\rm LOS}$ and temperature at $\log\tau\approx-4$. The $v_{\rm LOS}$--temperature phase distribution also clustered near $-90^\circ$, showing that the quarter-cycle phase relation is not specific to the line-core intensity diagnostic. 

Negative phase differences in Na--Ca $v_{\rm LOS}$ (Figure \ref{fig:NaCa_phase_coherence} A) indicate net upward propagation between the two atmospheric layers, while the apparent phase speeds are predominantly transonic, even when adjusting for the magnetic-field inclination (Figure \ref{fig:NaCa_phase_coherence} D). 
The limb-ward viewing geometry may reduce the Na--Ca coherence by sampling slightly different parts of inclined magnetic structures. For instance, for the inferred average height separation of $\Delta z$\,$\approx$\,325 km and $\mu=0.76$, the total projection offset is $\Delta r=\Delta z\tan[\cos^{-1}(\mu)]\approx278\,\mathrm{km}\approx0.38"$, with projected components $\sim$\,0.21"~along the slit and $\sim$0.32"~perpendicular to it. However, accounting for plausible along-slit shifts did not significantly increase the measured coherence between both velocity time series. This may indicate that the oscillations remain spatially coherent over scales comparable to the projected separation between the two formation heights. The sparse raster sampling prevents a comparable correction across the slit.
Compressive waves can also steepen into shocks, which are particularly strong in plage, as they propagate into decreasing density, generating non-sinusoidal waveforms and further reducing coherence. Conventional phase and coherence measurements may therefore not fully represent the underlying dynamics \citep[e.g.,][]{2008ApJ...683L.207R}. These effects do not contradict a slow-mode interpretation, but require that the \ion{Ca}{II} phase relationships be treated with caution. The \ion{Na}{I} D$_1$ phase relationships are more robust in that sense, as shock patterns are not observed in this line. 

At the \ion{Ca}{II} formation height, the phases among $I_{\rm core}$, $v_{\rm LOS}$, and $B_{\rm LOS}$ (Figure \ref{fig:internal_phase} B,D,F) are generally consistent with the slow standing or partially standing surface sausage-mode wave solutions of \citet{Moreels2013}. However, the magnetic phase distributions in \ion{Na}{I} D$_1$ (Figure \ref{fig:internal_phase} A,E) overlap qualitatively with the fast propagating surface sausage-mode wave solutions. Similarly to \ion{Ca} {II}, internally consistent phase relationships have been interpreted as longitudinal sausage and/or transverse kink waves, with the magnetic-velocity phase consistent with a superposition of upward and downward disturbances \citep{Fujimura2009}. Upward, downward and standing waves were also found to coexist along the spicules, and the downward component was attributed to reflection near the transition region \citep{2011ApJ...736L..24O}. Although their waves were transverse and sampled a different structure, the result demonstrates that an upward signal at lower heights can coexist with standing behavior higher in the atmosphere. A secondary fast or kink-like contribution cannot be excluded, since transverse displacement, field-inclination changes, or variations in magnetic filling factor can produce apparent $B_{\rm LOS}$ oscillations \citep{Ploner1997,2003ApJ...587..806M}. 

A plausible interpretation of finding signatures of both propagating and standing wave components is that compressive disturbances propagate upward through the lower chromosphere and undergo partial reflection at greater heights, producing a superposition of upward and downward components and therefore more locally standing-like phase relationships in \ion{Ca}{II}. Simulations with localized magnetic driving exciting coupled slow and fast waves provide a framework for such mixed signatures \citep{Riedl2021}. Approximately antiphased intensity-magnetic and quadrature velocity-magnetic behavior reported in plage \citep{Norton2021}, also show that similar phase combinations can arise from different mixtures of propagation, reflection, geometry, and radiative response. A quarter-cycle lag of photospheric magnetic perturbations relative to the Doppler velocity has also been reported in a plage region by \citet{Ji2021}, although the magnetic RMS were only around 1 G, possibly due to lower spatial resolution. 

Sausage modes have been identified most clearly in magnetic pores, where compact geometry and a well-defined intensity boundary allow direct measurements of area oscillations \citep[e.g.,][]{2011ApJ...729L..18M,2015A&A...579A..73M,2016ApJ...817...44F,2022ApJ...938..143G,2024A&A...688A...2J}. In particular, \citet{2015A&A...579A..73M} inferred slow sausage oscillations with a phase speed of approximately $5\,{\rm km\,s^{-1}}$, while \citet{2022ApJ...938..143G} found behavior consistent with partial reflection and standing-wave formation. Possible slow sausage-wave signatures have also been reported in isolated photospheric magnetic bright points through correlated intensity and area oscillations \citep{2021SoPh..296..184G}. \citet{2016ApJ...817...44F} measured transonic radial motions and magnetic perturbations of approximately $4$--$7\%$ in a background pore field, and several mode families have also been shown to carry appreciable fractions of wave energy within the same pore \citet{2024A&A...688A...2J}. The few-gauss perturbations we measured in plage correspond to about 5\% of a representative 200\,G field in the low chromosphere, comparable to the amplitudes reported in pores.  However, most pixels show only weak-to-moderate coherence in the $B_{\rm LOS}$ phase relationships. Sausage modes in plage are less well established, although tentative detections have been reported using photospheric magnetic and velocity diagnostics \citep{Fujimura2009,Norton2021} and chromospheric brightness and area oscillations in plage or enhanced-network bright features \citep{2023A&A...671A..69G}. 

Because plage magnetic elements are almost vertical through the photosphere and low chromosphere \citep[e.g.,][]{2020A&A...644A..43P}, the oblique viewing angle provides sensitivity to transverse perturbations that could arise from Alfvénic waves. However, the measured Doppler and magnetic RMS amplitudes imply a ratio of magnetic and kinetic energy densities ($\delta B_{\rm LOS}^2/\mu_0) /(\rho\delta v_{\rm LOS}^2)\approx80$ at the \ion{Na}{I} formation height and $\approx4$ at the \ion{Ca}{II} formation height. This imbalance is difficult to reconcile with the equipartition expected for simple linear Alfvén waves \citep{1942Natur.150..405A}. This does not rule out transverse waves, since equipartition need not hold in structured cylindrical flux tubes \citep[e.g.,][]{2013ApJ...768..191G}, and our LOS measurements do not recover the total wave energy integrated over the magnetic elements. 

In sunspots, wave-driven opacity changes can shift the effective formation height and alter the inferred $B_{\rm LOS}$ and $v_{\rm LOS}$ power \citep{2000ApJ...534..989B,2014ApJ...795....9F,2023A&A...670A.133F}. However, some observations contradict those findings \citep{2018A&A...619A..63J}. Opacity-driven variations in geometrical sampling height in pores might also contaminate inferred $B_{\rm LOS}$ oscillations at low frequencies \citep{2026FrASS..1314323S}. Under the usual assumption that the magnetic-field strength decreases with height, an opacity-induced shift of a simple absorption-line formation height would tend to produce in-phase variations of $I_{\rm core}$--$B_{\rm LOS}$, because upward shifts would sample both weaker field and lower line-core intensity \citep{Fujimura2009}. For \ion{Ca}{II}, shock-driven temperature perturbations may produce brighter emission at higher, weaker-field layers, creating an apparent anti-correlation between $I_{\rm core}$ and $B_{\rm LOS}$. Opacity effects in plage regions have yet to be quantified. We speculate that they may be less pronounced than in sunspots/pores because of systematically higher mass densities and temperatures. 

The inferred wave flux differences at different heights are significantly lower at the oblique viewing angle of these observations than in previous measurements obtained closer to the disk center \citep[e.g.,][]{Sobotka2016,2019ApJ...871..155R,Abbasvand2020b,Morosin2022,daSilvaSantos2024}. For a predominantly vertical, field-aligned velocity perturbation, projection effects alone could make the inferred flux about 42\% lower than at the disk center. However, some of the values reported in the literature are an order of magnitude higher than our average values. Differences in the probed height range and frequency bands make direct comparisons less clear. We note that these estimates account for only the acoustic wave flux. Although we cannot rule out the role these waves have in lower chromospheric heating, the multi-height measurements using the relative phases of intensity, velocity, and magnetic-field perturbations can constrain how wave modes evolve and magnetic configuration in plage regions changes with height. 

The comparatively large magnetic RMS amplitudes suggest that the associated magnetic-energy contribution may also be substantial, especially at the \ion{Na}{I} depth, even if part of the measured variability may arise from a temporal modulation of the line-formation height. Quantifying the associated Poynting flux would require the vector velocity and magnetic field. The stronger retained flux in plage is consistent with the magnetic channeling of low-frequency waves along inclined fields \citep{2014A&A...567A..62K,2019ApJ...871..155R}, while reduced power in the chromosphere around magnetic concentrations is also expected from magnetic shadows \citep[e.g.,][]{2007A&A...461L...1V}. In addition, radiative-MHD simulations suggest that vortex flows may further enhance slow-mode shock dissipation and localized heating along magnetic field lines \citep{2026arXiv260521230Y}. Identifying such vortex-related dynamics is challenging with ViSP slit scans, but high-cadence broadband imaging with DKIST/VBI could help test this scenario in future observations.

Using neutral \ion{Na}{I} and ionized \ion{Ca}{II} raises the possibility that partial ionization contributes to their differing responses in magnetized vs weakly magnetized regions. Ion-neutral collisions generally couple the species in the dense lower chromosphere, but finite drift and ambipolar diffusion can still modify magnetoacoustic propagation and dissipate magnetic perturbations as heat \citep[e.g.,][]{2016ApJ...819L..11S,2017PPCF...59a4038K}. Simulations indicate that wave energy can be dissipated through ion-neutral collisions without shock formation, although ion-neutral effects can enhance the dissipated energy in shock fronts \citep{2020A&A...635A..28W,MartinezSykora2020}. Ion-neutral effects are therefore a plausible contributor to the observed wave flux deposition. However, because the \ion{Na}{I} and \ion{Ca}{II} lines differ in formation height, their phase or amplitude differences cannot be interpreted as a direct measurement of ion-neutral decoupling. Isolating this process requires multi-fluid calculations and more nearly co-spatial neutral--ion line pairs. 

\section{Conclusions}

This work presents the first simultaneous two-height spectropolarimetric characterization of chromospheric oscillations in active-region plage using intensity, Doppler velocity, and magnetic-field measurements, supported by non-LTE inversion models. DKIST/ViSP observations of \ion{Na}{I} D$_1$\,5896\,\AA~and \ion{Ca}{II}\,8542\,\AA~allow us to track wave propagation from the low to middle chromosphere and relate it to the observed phase relationships, including those involving magnetic fluctuations, as well as to wave-energy estimates. Such measurements are less well established in plage than in other magnetic structures, such as sunspots and pores. They provide observational constraints on how MHD wave signatures and mode properties vary spatially and with height, and allow us to estimate wave energy dissipation rates in plage. This extends earlier studies that relied primarily on single-height diagnostics or lacked magnetic measurements at multiple heights, which are needed to more fully characterize the evolution of MHD wave modes through the atmosphere.

In plage, the internal $v_{\rm LOS}$--$I_{\rm core}$ phase relationships and the observed lag between \ion{Na}{I} and \ion{Ca}{II} are consistent with predominantly compressive oscillations propagating along the LOS with transonic phase speeds, characteristic of slow magnetoacoustic waves. Magnetic oscillations with RMS amplitudes of $\lesssim$10\,$\rm{G}$ accompany the intensity and velocity perturbations in plage , although they are detected more confidently in the \ion{Na}{I} line than in \ion{Ca}{II} line. This sets an upper limit on the magnitude of magnetic oscillations in the plage chromosphere to only a few gauss, which remains observationally challenging to detect.

A key result is that the magnetic phase relationships are not preserved between the two sampled heights. At the \ion{Ca}{II} height, the measured $I_{\rm core}$--$v_{\rm LOS}$--$B_{\rm LOS}$ phase relationships are broadly consistent with standing or partially standing slow sausage-like solutions, whereas the \ion{Na}{I} magnetic phases overlap qualitatively with propagating fast surface-sausage solutions. Kink modes may be secondary. However, a more definitive identification of magnetic flux-tube eigenmodes would require direct measurements of cross-sectional area variations, similar to those obtained in sunspots and small pores. Such measurements are more challenging in plage because of the lower contrast and smaller size of the magnetic elements. We therefore do not interpret these results as an unambiguous identification of distinct tube eigenmodes at different heights, particularly given additional effects such as opacity variations and line-of-sight projection. Nonetheless, the observations suggest that the observable MHD-wave signature evolves over only a few hundred kilometers in the chromosphere, consistent with a wave field that contains upward-propagating disturbances together with partial reflection and/or mode coupling at greater heights. This height-dependent behavior may, in turn, provide useful constraints on the atmospheric stratification and magnetic structure of plage regions.

Moreover, while our wave-flux calculations yield positive deposited fluxes, indicating net dissipation of the upward wave energy flux between the two sampled layers, they are substantially smaller than the expected chromospheric energy losses. Plage populations retain a larger fraction of their lower-layer flux than the non-plage populations. Thus, while these MHD disturbances seem to participate in energy transport, the measured kinetic component alone cannot account for the energy requirement. Quantifying the corresponding Poynting flux would require measurements of the vector magnetic field and velocity, which our data cannot constrain.

Future multi-line observations with higher sensitivity at different viewing angles are needed to improve measurements of the chromospheric magnetic vector and help distinguish intrinsic wave perturbations from LOS projection effects. Likewise, radiative-MHD or multi-fluid simulations are needed to separate the effects of reflection, shocks, ion-neutral processes, radiative losses, mode coupling, and magnetic geometry on the measured chromospheric energy deposition.

\section*{Conflict of Interest Statement}

The authors declare that the research was conducted in the absence of commercial or financial relationships that could be construed as a potential conflict of interest.

\section*{Funding}

AP was supported by grant PI~2102/1-1 from the Deutsche Forschungsgemeinschaft (DFG).

\section*{Acknowledgments}

The authors thank Dr. K. Reardon for his assistance with the line-core extractions and for introducing the proper orthogonal decomposition method, and Dr. T. Bogdan for discussions on wave energy estimation.
The research reported herein is based in part on data collected with the Daniel K. Inouye Solar Telescope (DKIST), a facility of the National Solar Observatory (NSO). NSO is managed by the Association of Universities for Research in Astronomy, Inc., and is funded by the National Science Foundation. DKIST is located on land of spiritual and cultural significance to Native Hawaiian people. The use of this important site to further scientific knowledge is done with appreciation and respect. This work used the Blanca condo computing resource at the University of Colorado Boulder. Blanca is jointly funded by computing users and the University of Colorado Boulder.

Taylor Sitterson also acknowledges his late father and grandfather, Allan E. Sitterson Jr. and Allan E. Sitterson Sr., whose encouragement and support in STEM allowed him to get this far.

\section*{Data Availability Statement}The SDO data can be obtained from the Joint Science Operations Center \url{http://jsoc.stanford.edu}. The DKIST datasets (ID: JPKFIO, GKIXRP, BWDSXN) can be found in the DKIST Data Center Archive \url{https://dkist.data.nso.edu/}. Derived analysis products and figure scripts will be available upon reasonable request.

\end{document}